\documentclass{article}

\usepackage{PRIMEarxiv}

\usepackage[utf8]{inputenc} 
\usepackage[T1]{fontenc}    
\usepackage{hyperref}       
\usepackage{url}            
\usepackage{booktabs}       
\usepackage{amsfonts}       
\usepackage{nicefrac}       
\usepackage{lipsum}
\usepackage{fancyhdr}       
\usepackage{graphicx}       
\graphicspath{{figure/}}     
\usepackage{amssymb}
\usepackage{amsthm}
\usepackage{amsmath}
\usepackage{bm}
\usepackage{subfig}
\usepackage{float}
\usepackage{booktabs}
\usepackage{diagbox}
\usepackage{multirow}
\usepackage{color}
\usepackage{CJK}
\usepackage{setspace}
\usepackage[section]{placeins}

\title{Development of a Hybrid Simulation and Experiment Test Platform for Dynamic Positioning Vessels
\thanks{\textit{\underline{Citation}}: 
\textbf{Changjun Hu, Quan Shi, Xin Li, Xiaoxian Guo. Development of a Hybrid Simulation and Experiment Test Platform for Dynamic Positioning Vessels. }} 
}

\author{
  Changjun Hu$^{1,2}$, Quan Shi$^{1,2}$,Xin Li$^{1,2}$,Xiaoxian Guo$^{*1,2}$ \\
  1 State Key Laboratory of Ocean Engineering, Shanghai Jiao Tong University , Shanghai,China\\
  2 SJTU Yazhou Bay Institute of Deepsea Technology, Sanya, China \\
  \texttt{\{Xiaoxian Guo\}xiaoxguo@sjtu.edu.cn} \\
}

\begin{document}
\maketitle

\begin{abstract}
The harsh ocean environment and complex operating condition require high dynamic positioning (DP) capability of offshore vessel. The design, development and performance evaluation of DP system are generally carried out by numerical simulations or scale model experiments. Compared with the time-consuming and laborious experiment, the simulation is convenient and low cost, but its results lack practical reference due to oversimplification of the model. Therefore, this paper presents a hybrid simulation and experiment test platform for DP vessels. Its characteristics are: the realistic calculation of environmental loads and motion response, the consistency of algorithms and parameters for simulation and experiment greatly shortening the time of experiment adjusting, switchable and online renewable controller facilitating algorithm testing. The test platform can test the performance of DP system and determine the operational time window. In the hydrodynamic simulation, the six degree-of-freedom model is used to describe the dynamic response of the DP vessel, considering the fluid memory effect and frequency-dependent hydrodynamic parameters. In the experiment, the similarity theory based on the same Froude number is used to ensure the consistency of control parameters with simulation. Finally, a case study of DP shuttle tanker is used to verify the credibility of the test platform.
\end{abstract}


\section{Introduction}
The DP control system can continuously activate the vessel’s propellers to balance the external disturbances (wind, waves and currents, etc.), and automatically control the position/heading of vessel in horizontal plane. Recently, with the deepening of marine development, the development of marine engineering technology is becoming more and more complex\cite{sorensen-SurveyDynamicPositioning-2011}. The DP technology has been vigorously promoted, because of its broad application prospects, such as drilling, pipe-laying, offlading, and diving support etc. However, DP system development and evaluation for ocean offshore vessel are highly complex and time-consuming.

    The performance of DP system should be tested in simulation and model experiment before commissioning test in real. And for determining offshore dynamic positioning operational sea state, numerical and experimental validations are also required. While the model test is always constrained by the significant consumption of time and money, time domain simulation is a common and convenient tool for design, analysis and predication of the DP system. Donnarumma et al. (2015) designed the DP controller structure by model-based design approach using simulation techniques\cite{donnarumma-Numericalmodelsship}. Tannuri et al. (2003) developed a computational simulator for DP Systems enabling the simulation of several DP operations, as drilling station keeping, pipe laying path following and those related to assisted offloading\cite{tannuri-DevelopmentDynamicPosition}. Martelli et al. (2022) designed DP system and evaluated its dynamic performance using a ship’s dynamic simulator\cite{martelli-TimedomainMethodologyAssess-2022a}. Zhan et al. (2013) developed a numerical station-keeping simulation in waves for a simplified FPSO with two DP systems\cite{zhan-DynamicPositioningSimulation-2013}. However, most of the above scholars use simple numerical models in simulation. These simple models only consider the linear superposition of low frequency ship maneuverability model and wave frequency model. These models ignore the effects of fluid memory and frequency dependent hydrodynamic parameters. So it is hard to capture the nonlinear response of the external exciting force on the structures. Simulation results of the platform motion, power consumption has no reliable guiding significance for engineering practice. As Fossen (2011) emphasized clearly, the simulation model should be able to reconstruct the motion response of the real physical system, where including the convincing environmental loads and the fluid-memory effects caused by the hydrodynamic coefficients\cite{fossenHANDBOOKMARINECRAFT}. Therefore, it is necessary to establish a DP simulator considering more accurate hydrodynamic environment simulation rather than just three degree-of-freedom (3 DOF) motion model using constant hydrodynamic parameters.

    In addition, the model experiment is an effective means to study the motion response of DP control for vessels. Based on the similarity theory, researchers have carried out a lot of research works on the scale model experiments of DP control. Loria et al. (2000) carried out a 1:70 scaled model ship to validete the separation principle for DP using noisy position measurements\cite{loria-SeparationPrincipleDynamic-2000}. Pettersen and Fossen carried out the experiment of underactuated DP of a ship using a model ship, scale 1:70\cite{pettersen-UnderactuatedDynamicPositioning-2000}. Tannuri et al. (2010) carried out the experiment of sliding mode control for a 1:150 scaled tanker\cite{tannuri-DynamicPositioningSystems-2010}. Hu et al. (2020) carried out a 1:37 scaled model DP experiment of a novel twin-lift decommissioning operation\cite{zhihuanhu-ExperimentalStudyLowspeed-2022}. A more common research and test method is the combination of numerical simulation and experiment method. Experimental tests were performed in combination with numerical analysis in order to validate the control algorithm. Leira et al. (2009) demonstrate the performance of their reliability-based DP system of surface vessels moored to the seabed both by numerical simulations and laboratory experiments on a model vessel\cite{leira-ReliabilitybasedDynamicPosition-2009}. Tannuri and Morishita carried out a simplified experiment composed of a scaled model to pre-validate their simulator of typical DP system\cite{tannuri-ExperimentalNumericalEvaluation-2006}. But it is worth noting that the experimental conditions and equipment are not so easy to construct and obtain, and the commissioning of experiments on site is also very complicated and difficult. Therefore, it is very meaningful to develop a hybrid simulation and experiment test platform, so reliable hydrodynamic simulation can be used to replace a part of experiment works to complete parameter adjustment in advance.

    This paper attempts to find a convenient, efficient and accurate test evaluation method for DP system. Therefore, more accurate numerical simulation, model experiment with parameter pre-adjustment, switchable algorithm and parameter control module are considered to build a hybrid simulation and experiment test platform. The simulation environment is constructed in combination with the hydrodynamic programs, including the calculation of frequency-dependent hydrodynamic coefficients and motion response considering fluid memory effect under environment loads of wind, waves and currents. Therefore, more accurate time domain simulation of dynamic positioning system motion response is realized. In addition, the experimental environment is constructed by a hardware framework using the scaled model of real vessel based on similarity theory of the same Froude number. During the experiment, all data are converted to the real ship scale to ensure the consistency of algorithm and control parameters in all numerical simulation, experiment and real ship. This consistency makes it possible to pre-adjust experimental parameters with simulation results. The DP controller ,equipped with switchable complete closed-loop control solution (i.e., reference filter, PID control, QP-based thrust allocation algorithm), has been developed to be compatible with both the simulation environment and the experiment environment.
    The present paper is organized as follows: in Section \ref{sec:Overall}, the overall structure and characteristics of hybrid simulation and experiment test platform are briefly introduced. In Section \ref{sec:Simulation}, the calculation of accurate hydrodynamic and motion response in numerical simulation is introduced. In Section \ref{sec:Experiment}, the experiment model scales 1:50 and hardware equipments such as thrusters and observer are shown, and the scale conversion used is introduced. In Section \ref{sec:Control}, we show a modular controller with switchable and online parameter adjustment functions. The results of simulation and experiment are summarized in Section \ref{sec:Results} and some concluding remarks are given at the end of the paper.

\section{Overall Structure of the hybrid platform}
    \label{sec:Overall}

\begin{figure}[ht]
        \centering
        \includegraphics[width=\textwidth]{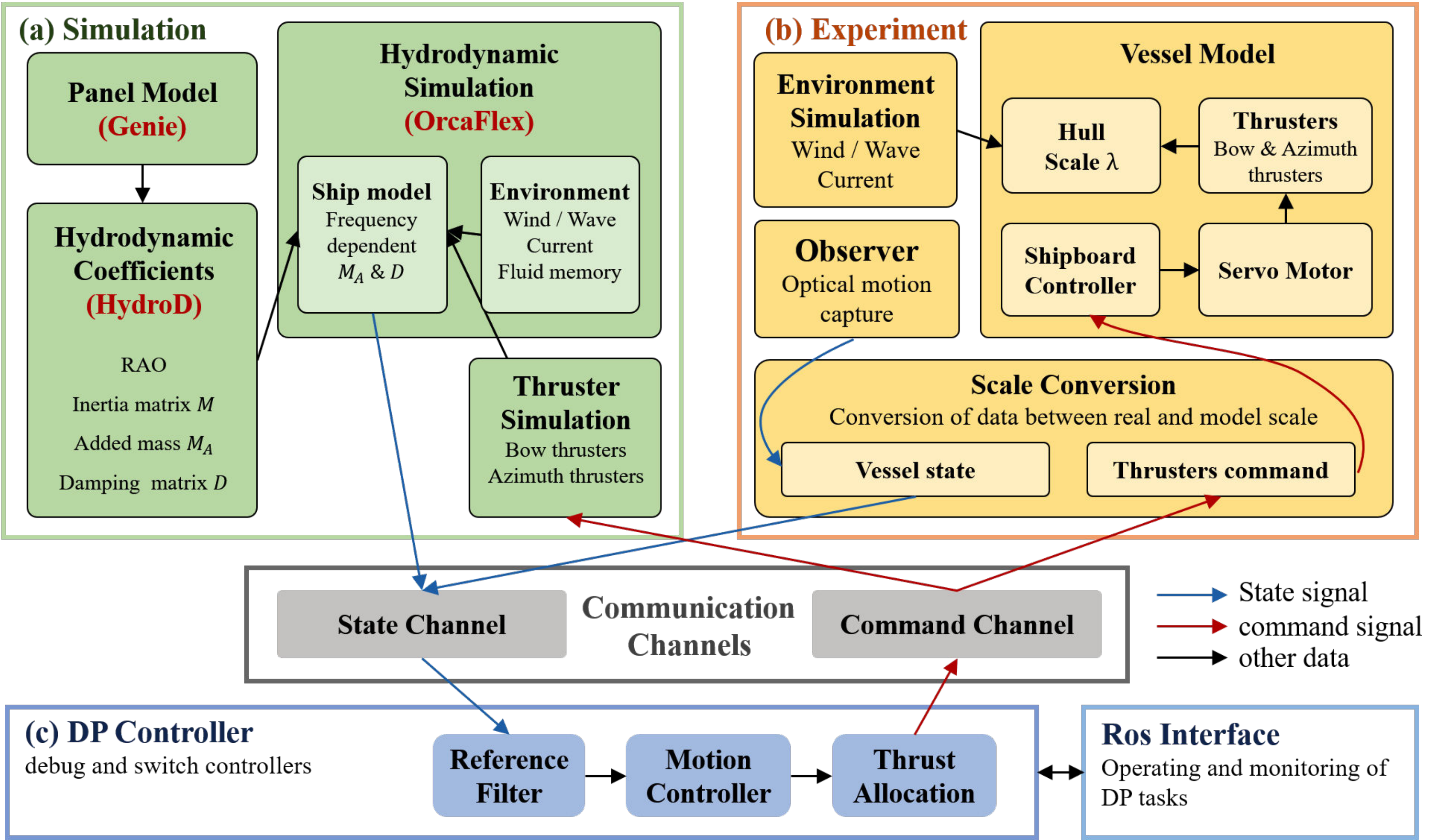}
        \caption{Block diagram of the comprehensive simulation and experiment platform}
        \label{fig:overall_structure}
    \end{figure}

    The framework of the hybrid simulation and experimental test platform mainly includes three parts: hydrodynamic simulation module, model experiment module and DP controller module. The block diagram of the hybrid test platform can be shown in Figure \ref{fig:overall_structure}. Among them, the hydrodynamics simulation module (a) using hydrodynamics calculation programs to compute the hydrodynamics, environmental loads, and the motion response of DP ship, details in section \ref{sec:Simulation} . The experiment module (b) means the scale model experiment carried out in the laboratory basin also used to test the performance of the DP system confirming with simulation, details in section \ref{sec:Experiment}.The control module (c) is implemented based on Robot Operating System (ROS) environment to meet the purpose of easy expansion and switchability, details in section \ref{sec:Control}. The signal interaction is realized through the local area network (LAN) TCP/IP communication protocol between controller and hydrodynamic simulation module or model experiment module, that is, receiving the ship's position/heading state and sending control commands.

    The design of this framework ensures the consistency of algorithms used in experiments and simulations. In order to achieve the same effective control effect in the scaled model experiment using the control parameters of the full scale obtained by simulation, more accurate calculation of hydrodynamic and motion response is realized in simulation, and the scale conversion is used in the model experiment to make the input and output data of each control loop be the full scale. Therefore, the pre-adjustment of control parameters can be completed in simulation, which can provide reference for model test parameters and shorten field adjusting time in experiment, tightly linking simulation and experiment together. It avoids the big difference of control parameters caused by different simulation and experiment scales mentioned by Ianagui et al.\cite{ianagui-RobustOutputFeedbackControl-2020}.

    The control module in the framework is modular designed and can be switchable, which is easy to expand, monitor and adjust parameters online. The realization of these characteristics is based on the use of ROS. ROS has become the standard platform approach for modern robots and is also used in the development of surface or underwater vehicles by other researchers \cite{henriklemckealfheim-DevelopmentDynamicPositioning-2018} \cite{chiaROSApproachMultimode2021}. The structure of ROS enables data to be transferred easily between modules through nodes and topics\cite{amandadattalo-ROSIntroductionROS-2018}.Topics are named buses over which nodes exchange messages. A node is a process that performs computation. Nodes are combined together into a graph and communicate with one another using streaming topics. In this paper, the modular development of the controller is realized , and each algorithm module can be switched independently. Online adjusting of parameters during program execution is also implemented, avoiding recompilation after each parameter adjustment, to improve the speed of parameter tuning.

\begin{figure}[ht]
        \centering
        \includegraphics[width=0.6\textwidth]{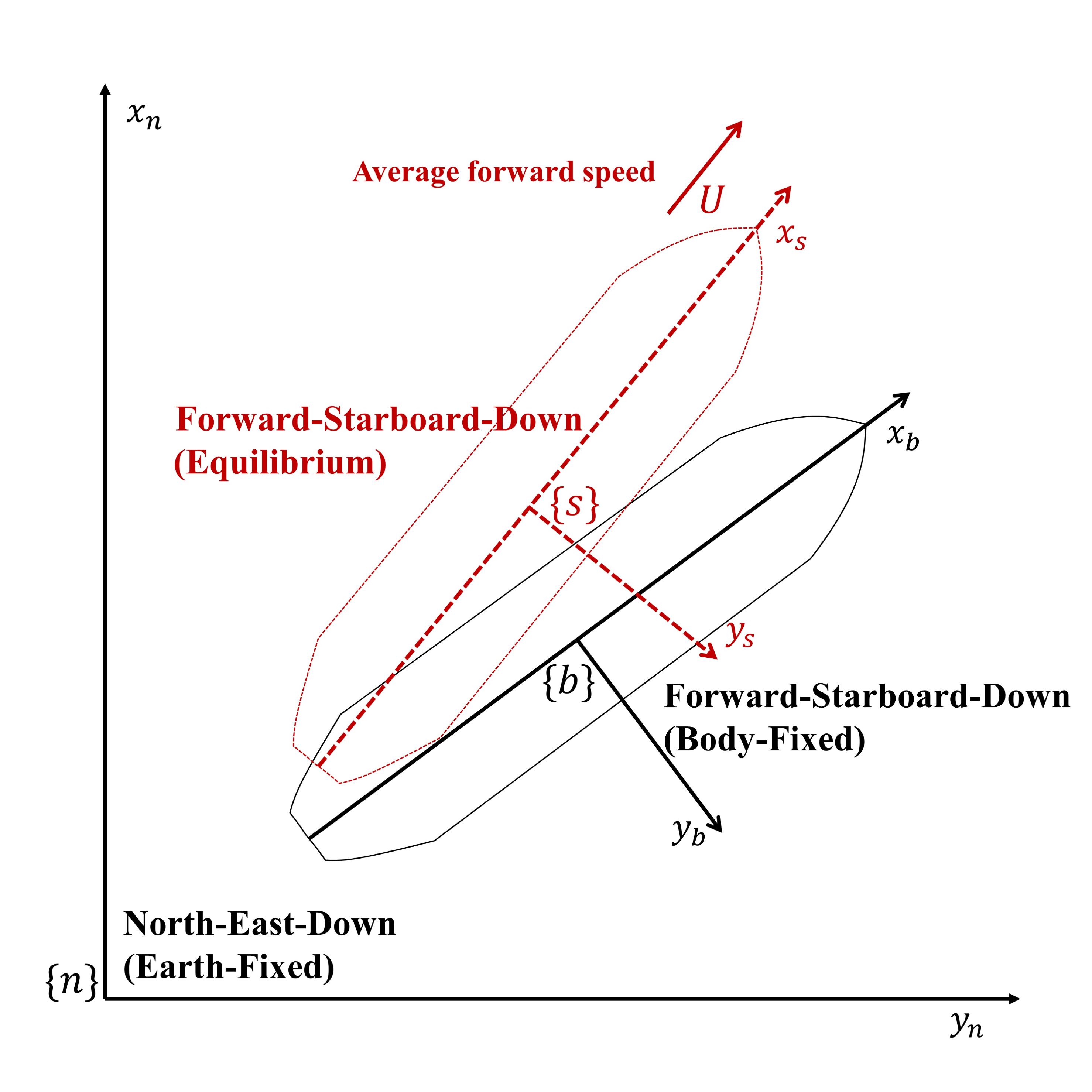}
        \caption{Coordinate system used in this paper}
        \label{fig:coordinate}
    \end{figure}

    \textbf{Coordinate system used in this paper} As shown in Figure \ref{fig:coordinate}, the \emph{North-East-Down} (NED) coordinate system $\{n\} = (x_n,y_n,z_n)$ is definited relative to earth as earth-fixed (global) reference frame, and the body-fixed (local) reference frame $\{b\} = (x_b, y_b, z_b)$ is a moving coordinate frame that is fixed to the vessel. The seakeeping reference frame $\{s\}=(x_s, y_s, z_s)$ is not fixed to the vessel; it is fixed to the equilibrium state of vessel. The {s} frame is considered inertial and therefore it is nonaccelerating and fixed in orientation with respect to the {n} frame, $U$ is the average forward speed.
    The position or velocities considered in this paper use the following representation: $\eta = [x, y, z, \phi, \theta, \psi]^T \in \{n\}$ is the vector of position/euler angles in earth-fixed reference frame; $ v = [u, v, w, p, q, r]^T \in \{b\}$ is the vector of velocities in body-fixed reference frame. $\delta \eta = [\delta x,\delta y,\delta z,\delta \phi,\delta \theta,\delta \psi]^T \in \{s\}$ is the vector of surge, sway and heave perturbations and  roll, pitch and yaw perturbations in seakeeping coordinates, and corresponding $\delta v = [\delta u, \delta v, \delta w, \delta p, \delta q, \delta r]^T \in \{s\}$ is perturbed velocities.

        \section{Simulation Moudule}
    \label{sec:Simulation}
    The structural block diagram of the hydrodynamics simulation system is shown in Figure \ref{fig:overall_structure}(a). In this module, the surface element model of wet surface is first established so that the Boundary Element Method (BEM) can be used to solve the hydrodynamic parameters. The boundary integral equations for diffraction/radiation potential on vessel surfaces are derived by Green function \cite{Newman-Algorithms-1985} and solved by a numerical method. The method discretes the integral equation on the average wet surface into several quadrilateral panels, and thus the velocity potential of each element can be calculated. Then, frequency-dependent hydrodynamics parameters can be obtained, such as additional mass, damping, and motion and load RAOs.

    Furthermore, the motion equation of vessel in time domain are obtained by applying Newton’s second law :
    \begin{equation}
        \begin{aligned}
            \dot{\eta}                                                                                    & =J(\Theta)v
            \\
            [M_{RB}+A(\infty)]\cdot \delta\dot{v}+\int_{0}^{\infty}{R(\tau)\delta v(t-\tau) d\tau}+ C\eta & ={F}_{\text{thruster}}+{F}_{\text{environment}}
            \\
            \delta{v} \approx v-\overline{v};  \ \overline{v}                                             & \approx U[1,-\delta \psi,\delta \theta,0,0,0]^T
            \\
            \delta\dot{v} \approx \dot v-\dot{\overline{v}};  \ \dot{\overline{v}}                        & \approx U[0,-\delta r,\delta q,0,0,0]^T
            \label{eq:motion}
        \end{aligned}
    \end{equation}
    where the $J(\Theta)$ is the transformation matrix between $\{b\}$ and $\{n\}$; $v$ represents the vector of velocity under $\{b\}$ ; $ \delta v, \ \delta\dot{v}$ are the perturbed velocities and accelerations under $\{s\}$; $\overline{v}, \ \dot{\overline{v}}$ represent the average velocities and accelerations; the mass matrix  $M_{RB}\in \mathbb{R}^{6 \times 6}$ represents the rigid inertia mass matrix and the added mass matrix $A(\infty)\in \mathbb{R}^{6 \times 6}$ is the infinite frequency added mass matrix giving the vessel's instantaneous response to acceleration; $\tau$ is a time lag integration variable; $R(\tau)$ is the a time-varying matrix of retardation functions proposed by Cummins \cite{Cummins-IRF-1962}, representing the memory effect of fluid; $C \in \mathbb{R}^{6 \times 6}$ is hydrostatic restoring force matrix. The right side of the equation represents the following external forces: $F_\text{thruster}$ is the vector of propulsion force and moment provided by the propellers, $F_\text{environment}\in \mathbb{R}^6$ is the vector of environment load including wave, wind and current loads.

    The retardation functions for $R(\tau)$ are calculated by Fourier integration of the frequency-dependent damping matrix $B(f)$ at frequency $f$, $R(\tau) = c(\tau)\int_{f=0}^{\infty}{[4B(f)\cos{(2\pi f t)}]df}$, where $c(\tau) = \exp{[-(3\tau / T_c)]}$ is a cutoff scaling function. $T_c$ is the cutoff time, after which the retardation function is truncated and assumed to be zero, because it decays to zero as $\tau \rightarrow \infty$ . \emph{Impulse Response Function} (IRF) is used to apply that retardation functions at each time step using a convolution integral to account for the past motion of the vessel. As opposed to the IRF, the infinite frequency added mass matrix $A(\infty)$ , giving the vessel's instantaneous response to acceleration, can be obtained from the retardation functions $R(s)$ and the added mass matrice $A(f)$ at frequency $f$ from the equation: $A(\infty) = A(f) + (1/{2\pi f})\int_{s=0}^{T_c}{R(s)\sin{(2\pi f s)}ds}$ \cite{Orcina2015}.

    The above motion response calculation is completed using the hydrodynamic software OrcaFlex. The Dynamic Link Library (DLL) of OrcaFlex and socket is used to communicate with the DP control system running in Ubuntu via TCP/IP communication protocol. This allows the two proccess to start at the same time and keep the same time step. Within the time loops the vessel position and heading state at each time step are sent to the DP controller which calculates the demand forces according to the errors from the set point, allocates the demand forces to the thrusters (called thrust command), and then sends the thrust command to the simulator. Simulator will calculate a new vessel state according to the thrust command and send the new state back to the DP controller, where the process repeats to achieve the time domain simulation for continuous dynamic positioning.

    \section{Experiment Moudule}
    \label{sec:Experiment}
\begin{figure}[ht]
        \centering
        \includegraphics[width=\textwidth]{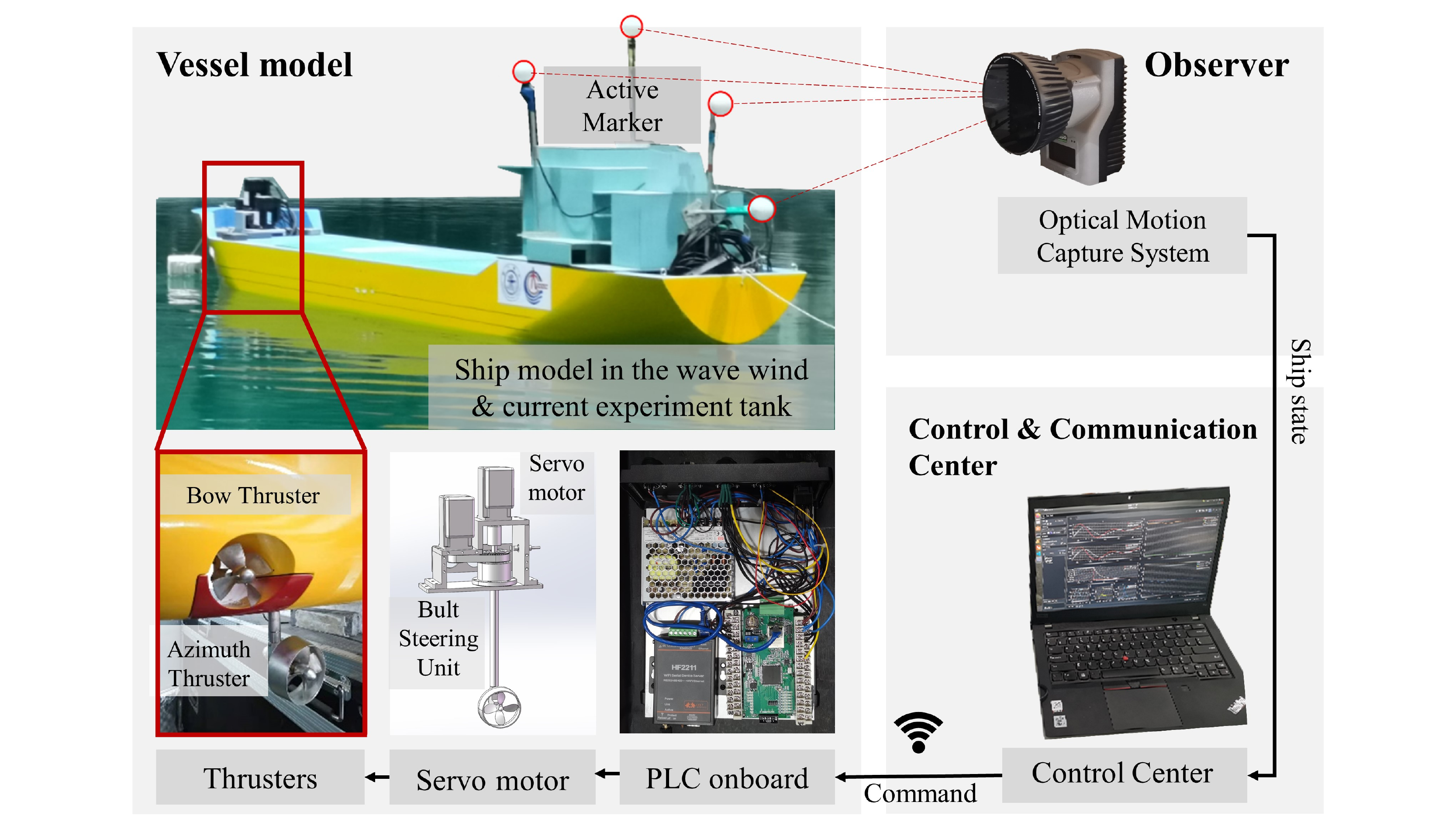}
        \caption{Block diagram of experiment platform}
        \label{fig:experiment_structure}
    \end{figure}

    In addition to simulation, the scaled model experiment is a convincible and reliable method to study the motion response of DP control for vessels. The model-scale experiments (1:50) were conducted in the deep-water offshore basin located at Shanghai Jiao Tong University, China. As shown in Figure \ref{fig:overall_structure}(b), the model experiment moudle is composed of scale conversion, vessel model, thrusters, servo motor, onboard controller, observer, etc. The details can be shown in Figure \ref{fig:experiment_structure}.

    \subsection*{ $\bullet$  Scale Conversion}

    In the experiment, the motion response of the reduced scale ship model is under environment and control loads in laboratory basin to simulate the real DP ship. For the model experiment in this paper, the data interacting with the DP program is converted to the real ship scale by assuming that the \textit{Froude number}
    \begin{equation}
        \text{Fn}= V/\sqrt{Lg} = constant
    \end{equation}
    where $V$ is velocity of the ship; $L$ is length of the ship; $g$ is the acceleration of gravity.The following scaling are obtained according to the scale ratio $\lambda$: length $L_s/L_m = \lambda$, angle $\phi_s/\phi_m = 1$, linear velocity $V_s/V_m = \sqrt{\lambda}$, angular velocity $r_s/r_m = \sqrt{1/\lambda}$, density of water $\rho_s/\rho_m = \gamma$, force $F_s/F_m = \gamma \lambda^3$, moment $N_s/N_m =\gamma \lambda^4$, time $T_s/T_m = \sqrt{\lambda}$.

    In each control loop, the position state of model scale obtained by observer will be converted to full scale and then sent to the control center. Similarly, the full scale thrust commands issued by the controller will also be converted to model scale and then sent to PLC and acted on the hull.Therefore, the control center always processes full-scale data to ensure the consistency of algorithm and control parameters in numerical simulation, experiment and real ship.

    \subsection*{ $\bullet$  Vessel model}

\begin{figure}[ht]
        \centering
        \includegraphics[width=4in]{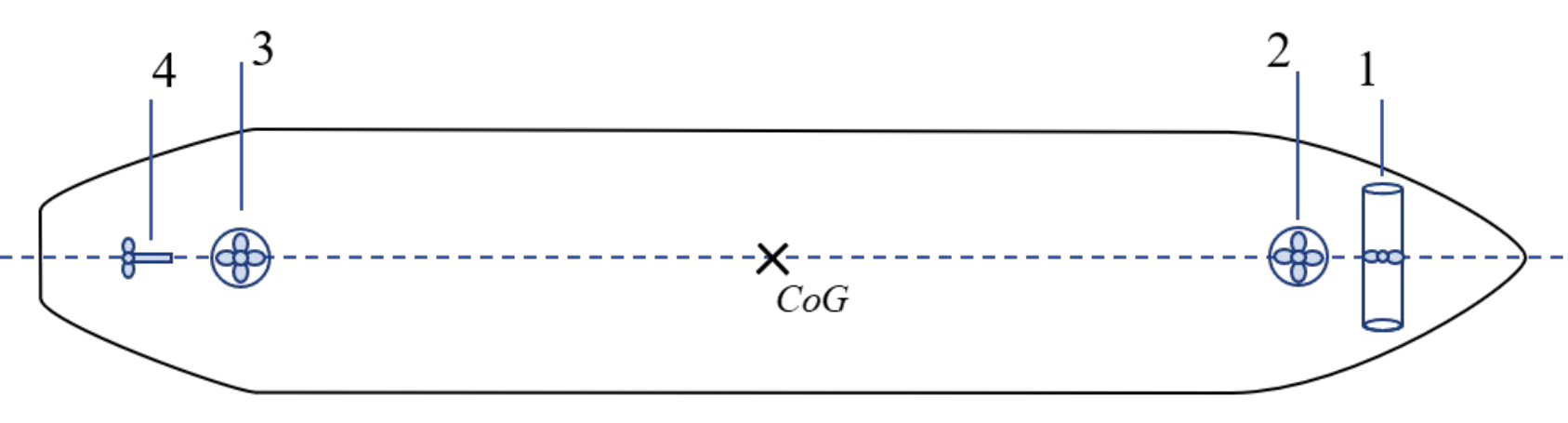}
        \caption{Arrangement of thrusters}
        \label{fig:thrusters}
    \end{figure}

    The experimental vessel model used in this paper is a model of a shuttle tanker (ST) scale $\lambda=1:50$. The principal dimensions of the ST is shown in Table \ref{tab:main_dimension}. The ST is equipped with four thrusters, two of which are azimuth thrusters, a lateral thruster at the bow and a main thruster at the stern. The arrangement of thrusters and their location and performance parameters are shown in Figure \ref{fig:thrusters} and Table \ref{tab:location_thrusters}.

    \begin{table}
        \renewcommand{\arraystretch}{1.2}
        \caption{Main Dimension of ST ($\lambda$=1:50)}
        \label{tab:main_dimension}
        \centering
        \setlength{\tabcolsep}{3mm}{
            \begin{tabular}{l c c c}
                \toprule 
                Main Dimension                & Unit & Full Scale          & Model Scale \\
                \midrule 
                Length Overall (Hull)         & $m$  & 137                 & 2.74        \\
                Length between Perpendiculars & $m$  & 134.6               & 2.692       \\
                Breadth Moulded               & $m$  & 22.8                & 0.456       \\
                Depth Moulded to Main Deck    & $m$  & 12.5                & 0.25        \\
                Draught                       & $m$  & 8.655               & 0.1731      \\
                Displacement                  & $kg$ & 21402$\times$10$^3$ & 167.04      \\
                \bottomrule 
            \end{tabular}}
    \end{table}

    \begin{table}
        \renewcommand{\arraystretch}{1.2}
        \caption{Location (relative to gravity center) \& capacity of thrusters}
        \label{tab:location_thrusters}
        \centering
        \setlength{\tabcolsep}{8mm}{
            \begin{tabular}{c c c c}
                \toprule 
                Thruster & X[m]   & Y[m] & Max Thrust[KN] \\
                \midrule 
                No.1     & 60.59  & 0    & 246            \\
                No.2     & 57.00  & 0    & 275            \\
                No.3     & -40.34 & 0    & 275            \\
                No.4     & -61.17 & 0    & 480            \\
                \bottomrule 
            \end{tabular}}
    \end{table}

    Those thrusters driven by servo motors, and a belt steering unit and a additional motor are used to adjust the direction of each azimuth thruster. The velocity of servo motors are controlled by the Programmable Logic Controller (PLC) running onboard. Since the shuttle tanker is equipped with 2 sets of azimuth thrusters and 1 set of lateral thruster and 1 set of main thruster, we require 6 sets of servo motors to drive these thrusters. Therefore, this ship model uses a 48V DC battery pack to power the servo motors and the PLC simultaneously.

    \subsection*{ $\bullet$  Control and Communication Center}

    The main body of the control center is a computer installed on shore running DP programs, receiving ship state and giving thrust commands. The control center wired connectes to observer for receiving ship state. The industrial WLAN is used for communication between PLC and control center, consist of one Accessing Point (AP) onshore connected the control center and one onboard client module running in PLC. PLC receives the thrusters command and feedbacks the actual speed information to the control center through the WLAN.

    \subsection*{ $\bullet$  Observer}

    The use of optical motion capture system, makes it possible to measure the position of the model in a non-contact manner, using the Qualisys Track Manager (QTM) software. The position and speed informations of the model obtained by the optical motion capture system are sent to the control center running the DP control algorithm. After receiving the model state, the control center outputs the thruster command after calculation, thus forming a closed-loop control loop.

    \section{Control Module}
    \label{sec:Control}

\begin{figure}[ht]
        \centering
        \includegraphics[width=\textwidth]{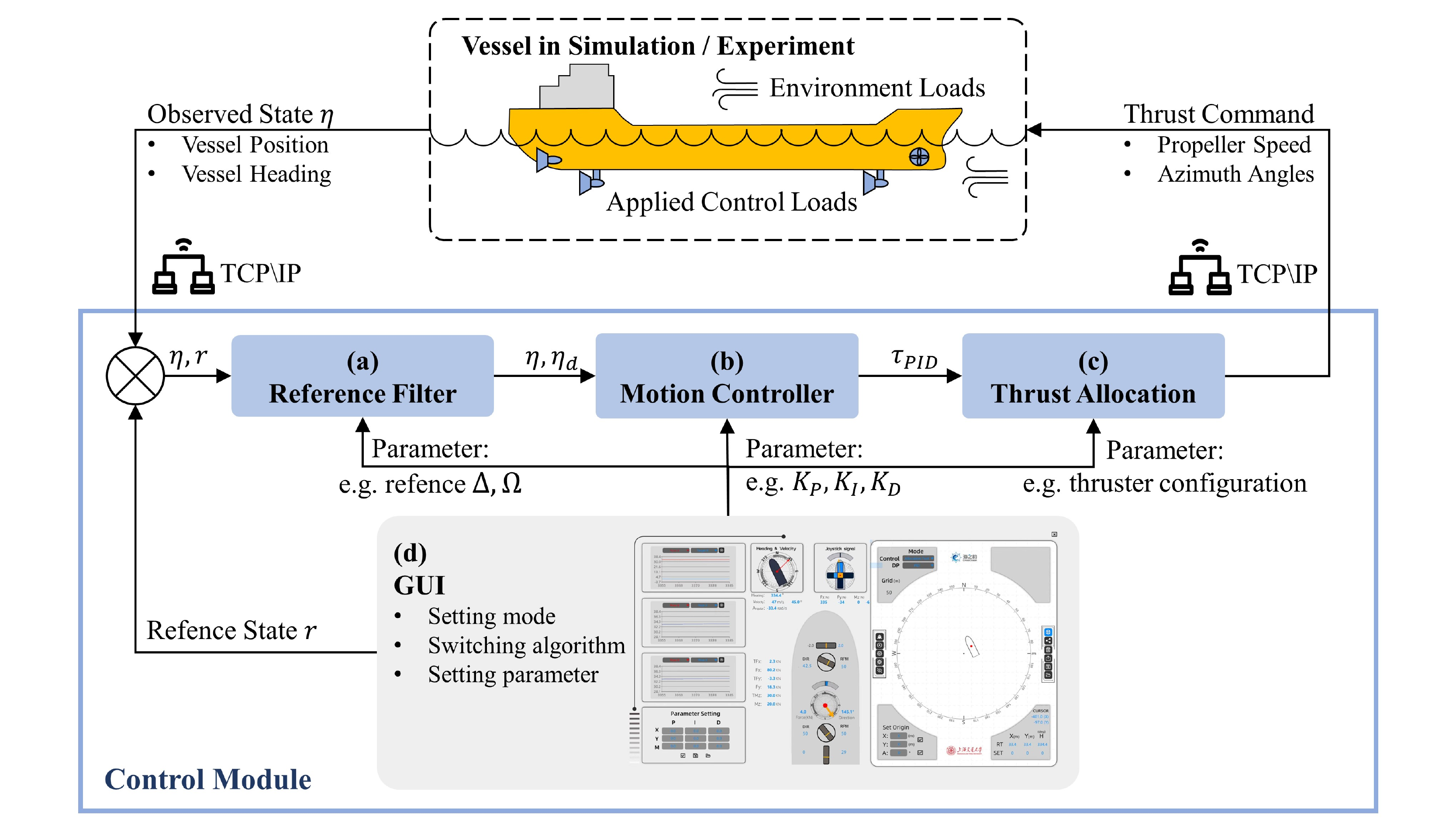}
        \caption{Diagram of DP control system}
        \label{fig:ros_node}
    \end{figure}

    The control module algorithm is based on Robot Operation System (ROS) environment.The diagram of this control system is shown in Figure \ref{fig:ros_node}. In order to realize the efficient test and verification of the DP algorithm, the loose coupling modular design of the algorithm is needed. Modular design allows algorithms with different functions to be independent of each other, and different algorithms with the same function are switchable and can be replaced by each other. Therefore, ROS is chooesn as the basis of the control system, which can provide an operating system-like capability for creating modular systems using nodes. In addition, in order to achieve online parameter adjusting, we need to be able to real-time revalue some variables (e.g. $K_P , K_I, K_D$ in PID controller) during program execution. The ROS parameter server helps us achieve this, because its parameter server provides a shared, multi-variate dictionary that is accessible via network APIs. Nodes use this server to store and retrieve parameters at runtime, avoiding recompilation after each parameter adjustment \cite{bradmillerParameterServerROS2018}.

    As shown in Figure \ref{fig:ros_node}, the TCP/IP communication protocol is used to implement data exchange between control modules and simulation or experiment. In each control closed-loop time step, the control module receives position and speed information of ship model from simulator running on WIN system during numerical simulation, or from optical motion capture system during vessel model experiment. After the calculation is completed, the control module sends thrust commands to simulator software during numerical simulation, or to onboard PLC during experiment. Note that all the state data sent to the control module and the thrust commands given by the controller are based on the full scale ($\lambda$=1:1), and thrust commands is applied to the model test after scaling using the similarity theory according to the corresponding scale, in order to ensure the consistency of the control algorithm applied to simulation and experiments in this way.

    The nodes named (a)Reference Filter, (b)Motion Controller and (c)Thrust Allocation algorithm in the blue block in the Figure \ref{fig:ros_node} are the core parts of the control module. Algorithms used in all those modules can be easily switched by algorithms with the same function. Algorithms used for examples in this paper use the corresponding theory shown in the fellowing.

    \subsection*{ $\bullet$ Reference Filter}

    In order to plan a feasible reference path between the ship and the setpoint, the physical limitations of the ship's velocity and acceleration need to be fully considered. In order to process the step signal of the reference target, Fossen et al. (2011) constructed a ship position and heading reference model \cite{fossenHANDBOOKMARINECRAFT} by connecting the first-order low-pass filter and mass damping spring system in series, which is used to generate smooth reference path.
    \begin{equation}
        \label{eq:reference}
        \eta_{d}^{(3)}+(2\Delta +I)\Omega {{\ddot{\eta }}}_{d}+(2\Delta +I){{\Omega }}^{2}{{\dot{\eta }}}_{d}+{{\Omega }^{3}}{{\eta }_{d}}={{\Omega }^{3}}{{r}}
    \end{equation}
    where $\eta_{d}= [x_d, y_d, \psi_d]$ is desired position and heading vector calculated with Eq. \ref{eq:reference}; $ r = [x_r, y_r, \psi_r]$ is setpoint  inputed by operators; $\Delta$ and $\Omega$ are diagonal matrices composed of relative damping coefficients and natural frequencies, respectively.

    \subsection*{ $\bullet$ Motion Controller}
    \label{par:PID}

    The PID controller is a robust and industry standard used in many applications. As the goal of this paper is to validate the test platform, then classical PID controller is a solid choice. The PID controller attempts to minimise the error $e(k)$ as the difference between a desired set point $\eta_d$  and an observed state $\eta_o = [x, y, \psi]$ in the body-frame coordinate system. The error can be expressed as:
    \begin{equation}
        e(k)=\eta_o(k)-{{\eta}_{d}}(k)
    \end{equation}

    In order to realize PID controller on computer, the equation is discretized and the following equation is obtained:
    \begin{equation}
        {\tau }_{PID}(k)={K}_{P}e(k)+{K}_{I}\sum\limits_{i}^{k}{e(i)}h+{K}_{D}\frac{1}{h}(e(k)-e(k-1))
    \end{equation}
    where ${K}_{P},{K}_{I},{K}_{D}$ are the proportional, integral and derivative parameters respectively; $h$ is the time step. When tuning the controller manually, operators can think about the physical meaning of parameters to determine the parameters. Under the above formula definition, the corresponding units are:
    \begin{equation}
        {\tau }_{PID}:\left[\begin{matrix}kN \\ kN \\ kNm \\ \end{matrix}\right],
        {K}_{P}:\left[\begin{matrix}\frac{kN}{m} \\ \frac{kN}{m} \\ \frac{kNm}{rad} \\ \end{matrix}\right],
        {K}_{I}:\left[\begin{matrix}\frac{kN}{m\cdot s} \\ \frac{kN}{m\cdot s} \\ \frac{kNm}{rad\cdot s} \\ \end{matrix}\right],
        {K}_{D}:\left[\begin{matrix}\frac{kN}{m/s} \\ \frac{kN}{m/s} \\\frac{kNm}{rad/s} \\ \end{matrix}\right]
    \end{equation}
    A frequently used method of tuning the PID is increasing the proportional gain until oscillations around the setpoint is achieved. Then adding derivative gain functioning as dampening to remove the oscillations. Lastly adding integral action removes the constant offset \cite{henriklemckealfheim-DevelopmentDynamicPositioning-2018}. During the adjustment of the control parameters on 3 DOF, maintaining the desired heading angle takes priority over position to produce the desired results \cite{shiCompositeFinitetimeAdaptive2022a}.

    \subsection*{ $\bullet$ Thrust allocation}

    The role of thrust distribution is to allocate the total desired control force and moment calculated by the motion controller to each actuator, so that the force and moment applied by the actuator on the ship is the same as the desired control force. The force acting on a ship can be written as:
    \begin{equation}
        {{\tau }_{thruster}}\text{=}B(\alpha )\cdot u
    \end{equation}
    \begin{equation}
        {{B}_{3\times m}}(\alpha )=\left[
            \begin{matrix}
                \cos {{\alpha }_{1}}                       & \cdots & \cos {{\alpha }_{m}}                       \\
                \sin {{\alpha }_{1}}                       & \cdots & \sin {{\alpha }_{m}}                       \\
                -l_{y1}\cos{\alpha_1}+l_{x1}\sin{\alpha_1} & \cdots & -l_{ym}\cos{\alpha_m}+l_{xm}\sin{\alpha_m} \\
            \end{matrix}
            \right]
    \end{equation}
    where $u = (u_1, u_2, \cdots, u_m)^T \in \mathbb{R}^m $ represents the thrust of each propeller;  $\alpha = (\alpha_1, \alpha_2, \cdots, \alpha_m)^T \in \mathbb{R}^m $  indicates the thrust direction of propeller $i$; $l_{xi}$and $l_{yi}$ represent the X and Y coordinate of propeller $i$  in the body-frame coordinate system, respectively.

    For thrust distribution problems, power consumption, angle change, speed change, etc., are frequently taken as optimization objectives to minimize the cost of generating thrust, and quadratic programming (QP) problems with linear constraints are constructed \cite{johansen-ConstrainedNonlinearControl-2004}. The QP problems can be solved very reliably and efficiently \cite{Convex-optimization-2004}. Singularity avoidance is also considered, see \cite{johansenConstrainedNonlinearControl2004a} for the details. The optimization objective function constructed is as follows:
    \begin{equation}
        J(\alpha , u, s)=\sum\limits_{i=1}^{m}{{{W}_{i}}({{u}_{i}})}+{{(\alpha -{{\alpha }_{0}})}^{T}}\Omega (\alpha -{{\alpha }_{0}})+{{s}^{T}}Qs +\frac{\rho }{\varepsilon +det(B(\alpha ){{B}^{T}}(\alpha ))}
    \end{equation}
    where  ${{W}_{i}}({{u}_{i}}) = {\mid u_i \mid}^{3/2} $ is the power of propeller $i$; $(\alpha-\alpha_0)$ is the propeller angle variation between the next time step and the present moment in operation; $s$ is the relaxation factor term in thrust distribution, that is, the allowable resultant error; and $Q$ is the weight of the relaxation factor. The last item is set to avoid singularity of $B(\alpha)$, the $\rho>0$ is weighted parameter, the $\varepsilon$ is a small positive definite parameter.

    In solving the thrust allocation algorithm, the thruster angle, max thrust and change rate should be constrained. The linear constraint conditions are as follows:
    \begin{equation}
        \begin{aligned}
             & u_{\min}\le u \le u_{\max}                                        \\
             & \Delta u_{\min} \le u-u_0 \le \Delta u_{\max}                     \\
             & \alpha_{\min}\le \alpha \le \alpha_{\max}                         \\
             & \Delta \alpha_{\min} \le \alpha-\alpha_0 \le \Delta \alpha_{\max} \\
        \end{aligned}
    \end{equation}

    \subsection*{ $\bullet$ Graphical User Interface (GUI)}

    The (d) GUI was programed using the QT application, integrated with ROS. The interface allows operators to control vessel in either the simulation or experiment in same way. making the online parameter adjustment in the system more convenient, such as the target state, controller parameter revalue or control algorithm switching.

    \section{Simulation and experiment results}
    \label{sec:Results}

    \subsection{Position keeping test}
    \label{sec:Results:pk}
\begin{figure}[ht]
        \centering
        \subfloat[{Maximum offset [m]}]{\includegraphics[width=0.4\textwidth]{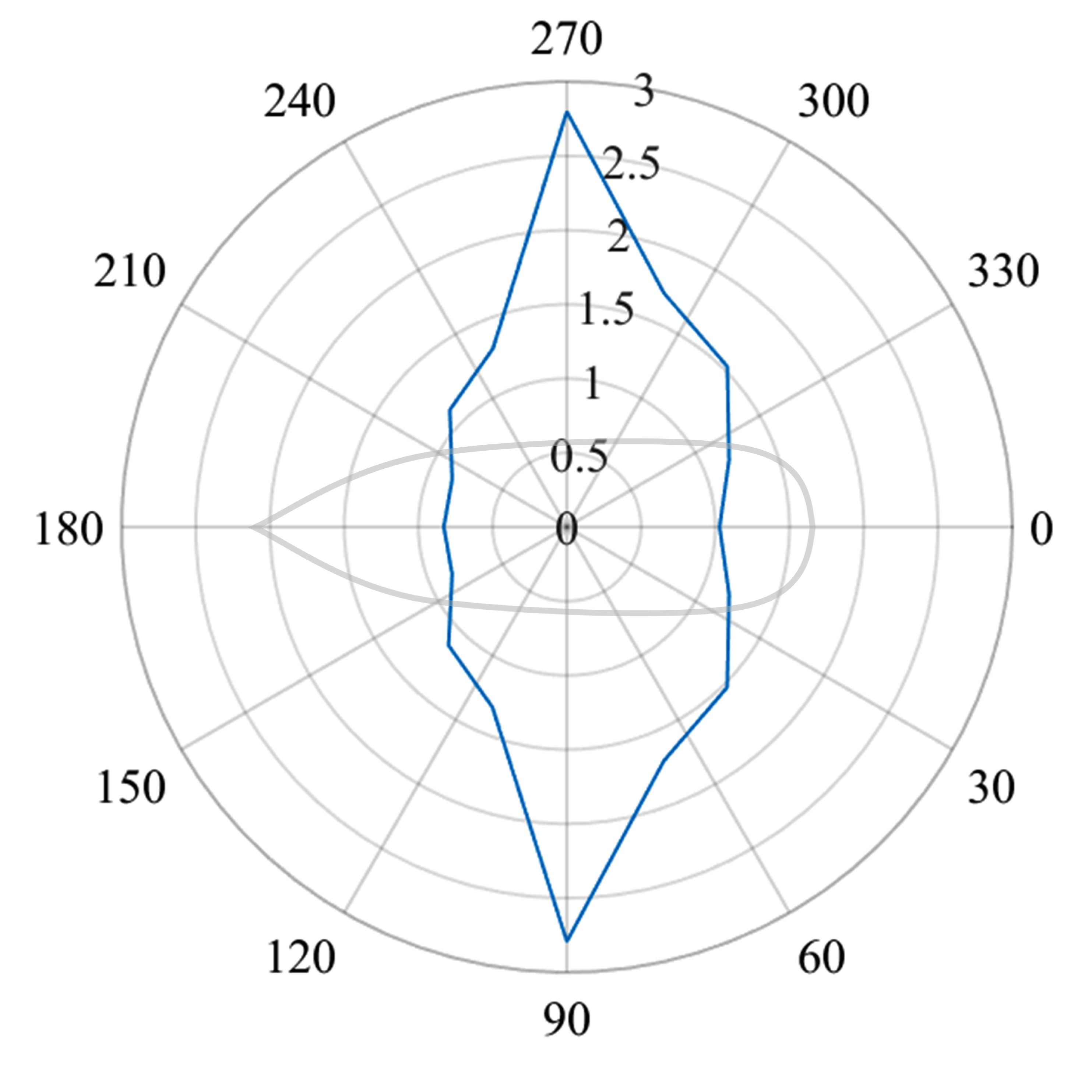}}
        \subfloat[{Maximum heading offset [$^{\circ}$]}]{\includegraphics[width=0.4\textwidth]{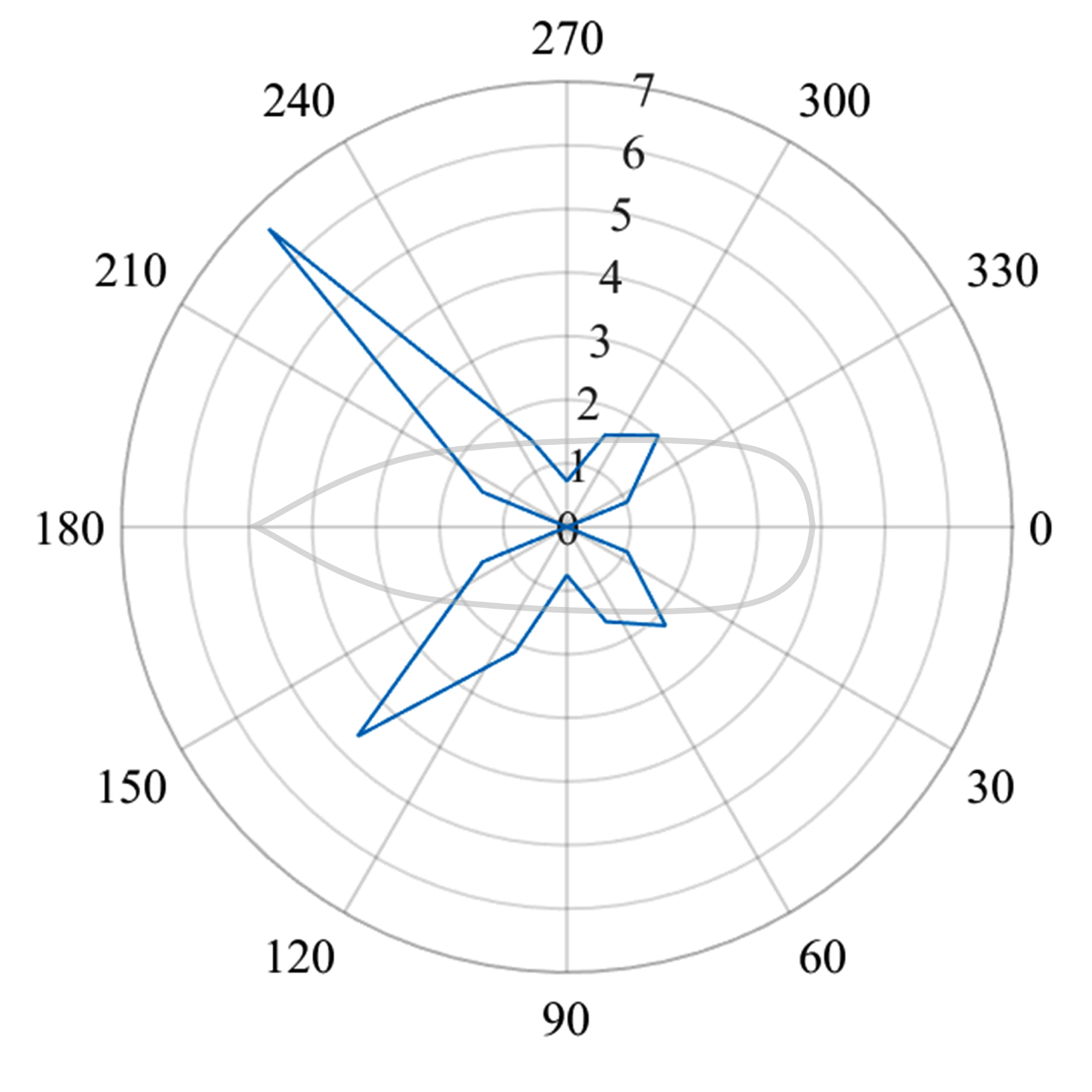}}
        \caption{Maximum offset of the ST}
        \label{fig:max_offset}
    \end{figure}

    Benefiting from the design of algorithm consistency in this paper, we can first adjust the control parameters in the simulation and provide reference for the experiment. In order to tuning the parameters and verify the control ability of the control module, the numerical simulation test of position keeping under different environmental loads were carried out. During simulation the DP ST is under the environmental condition assuming that wave ($Tp = 8s, Hs = 1.5m$),wind ($v_{wind}=5m/s$) and current ($v_{current}=0.5m/s$) loads are applied in the same direction. In the begining of the test, referring to the tuning method of PID parameters mentioned in section \ref{par:PID}, the PID parameters with good performance are obtained by using the convenient online parameter adjustment technology, show in Table \ref{tab:PID_gains}. Then a series of tests were conducted to obtain those DP capability plots of the ST for maximum offset, heading angle, mean thrust under environment load from 0$\sim$360$^\circ$ presented in Figure \ref{fig:max_offset}.

    \begin{table}
        \renewcommand{\arraystretch}{1.2}
        \caption{The PID gains.}
        \label{tab:PID_gains}
        \centering
        \setlength{\tabcolsep}{3mm}{
            \begin{tabular}{c c c c c c c c c}
                \toprule 
                \multicolumn{3}{c}{$K_P$} & \multicolumn{3}{c}{$K_I$} & \multicolumn{3}{c}{$K_D$}                                               \\
                X                         & Y                         & $\Psi$                    & X    & Y    & $\Psi$ & X    & Y    & $\Psi$ \\
                \midrule 
                400                       & 400                       & 400000                    & 0.05 & 0.05 & 2      & 2000 & 2000 & 200000 \\
                \bottomrule 
            \end{tabular}}
    \end{table}

    \FloatBarrier
    It can be clearly seen that the lateral environmental load has the greatest influence on the positioning of the ship, and the oblique environmental load has the greatest influence on the heading control.

\begin{figure}[ht]
        \centering
        \subfloat[{Mean Thrust X [KN]}]{\includegraphics[width=0.3\textwidth]{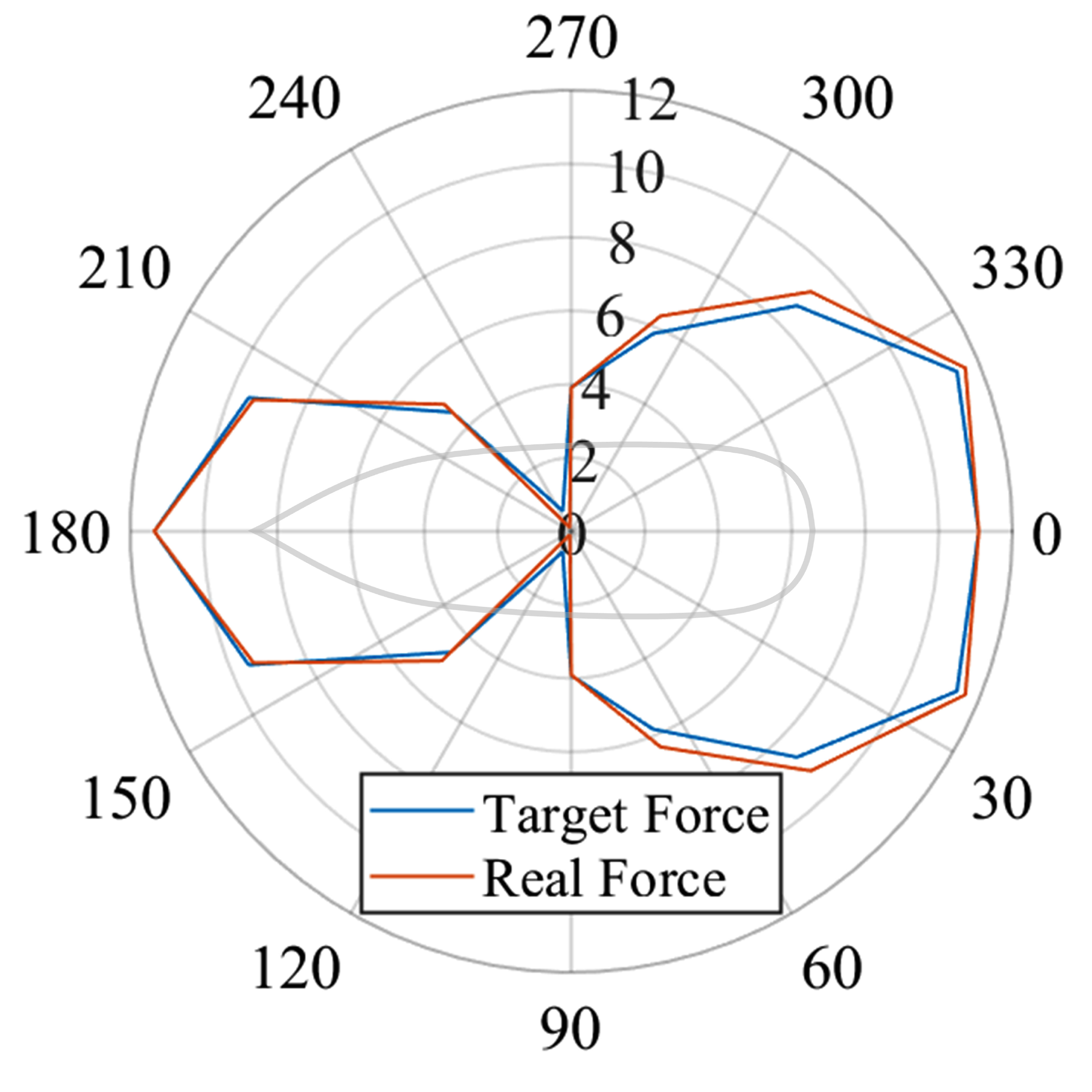}}
        \subfloat[{Mean Thrust Y [KN]}]{\includegraphics[width=0.3\textwidth]{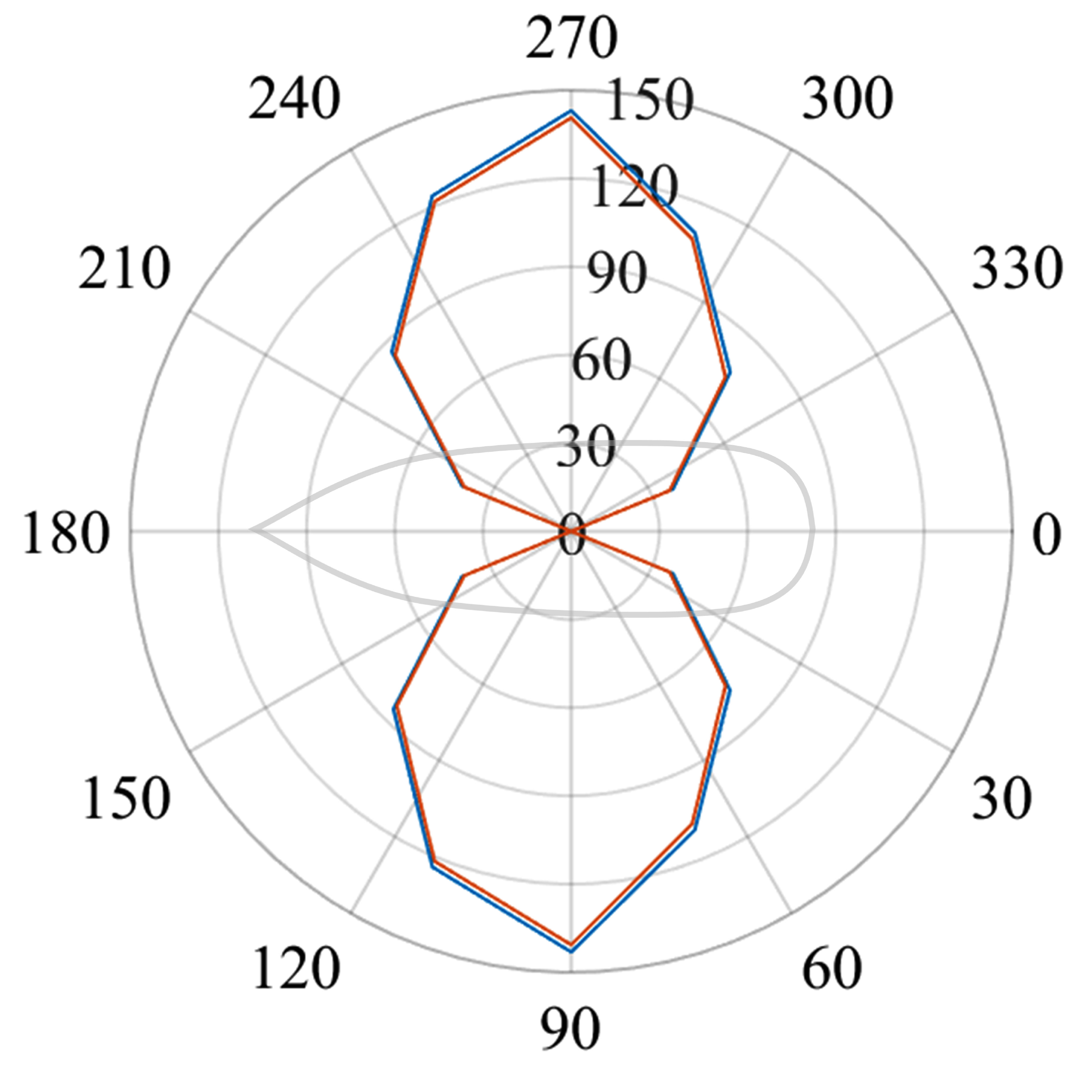}}
        \subfloat[{Mean Moment N [KN$\cdot$m]}]{\includegraphics[width=0.3\textwidth]{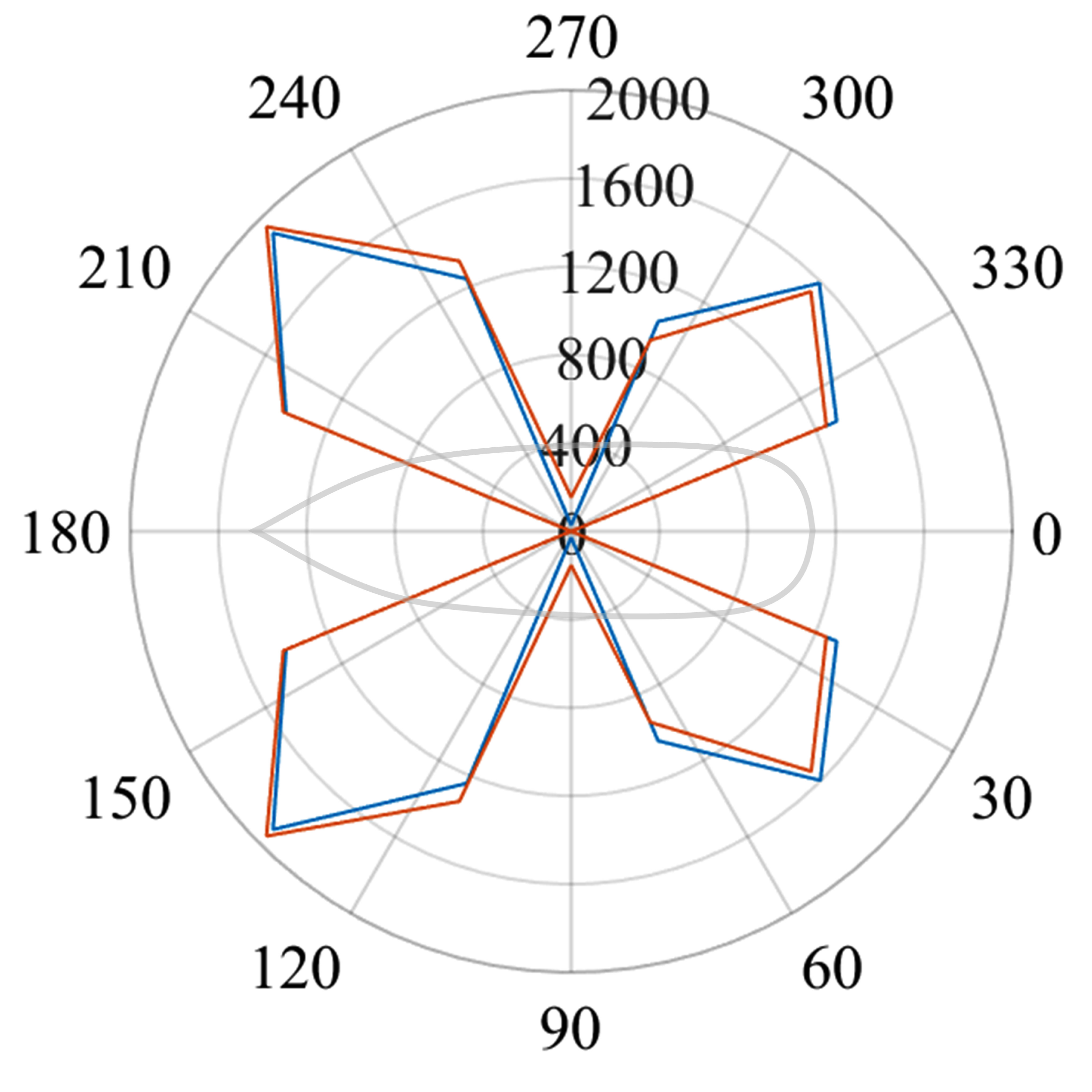}}
        \caption{Mean control force of the ST}
        \label{fig:mean_ctrl_force}
    \end{figure}

    This is also illustrated by the control forces given by the controller under the action of environmental forces in different directions shown in Figure \ref{fig:mean_ctrl_force}. The controller gives a larger lateral thrust to balance the lateral environmental forces and a larger moment to resist the moment brought by the oblique environmental forces.

    \FloatBarrier

    \begin{figure}[ht]
        \centering
        \includegraphics[width=0.9\textwidth]{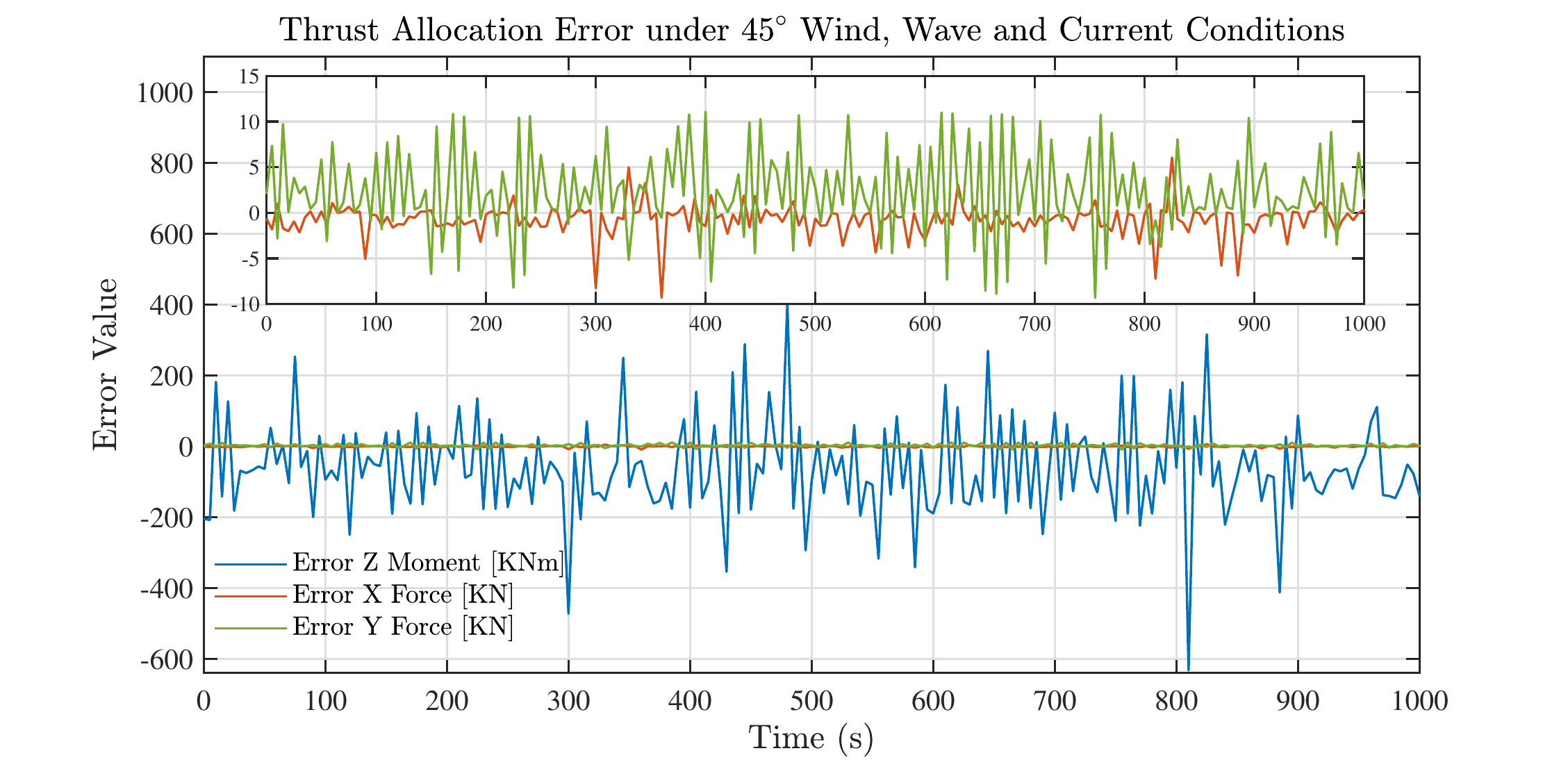}
        \caption{Thrust allocation error under Env.1 of 45$^{\circ}$}
        \label{fig:thrust_error}
    \end{figure}
    \FloatBarrier

    A specific time histories of thrust error is used to demonstrate the performance of the thrust allocation algorithm. The thrust allocation error under the environment force in 45 degree direction is shown in Figure \ref{fig:thrust_error}, and it can be seen that the error of thrust in X and Y directions is within 5\%, and the error of torque is within 10\%.

\begin{figure*}[ht]
        \centering
        \includegraphics[width=0.9\textwidth]{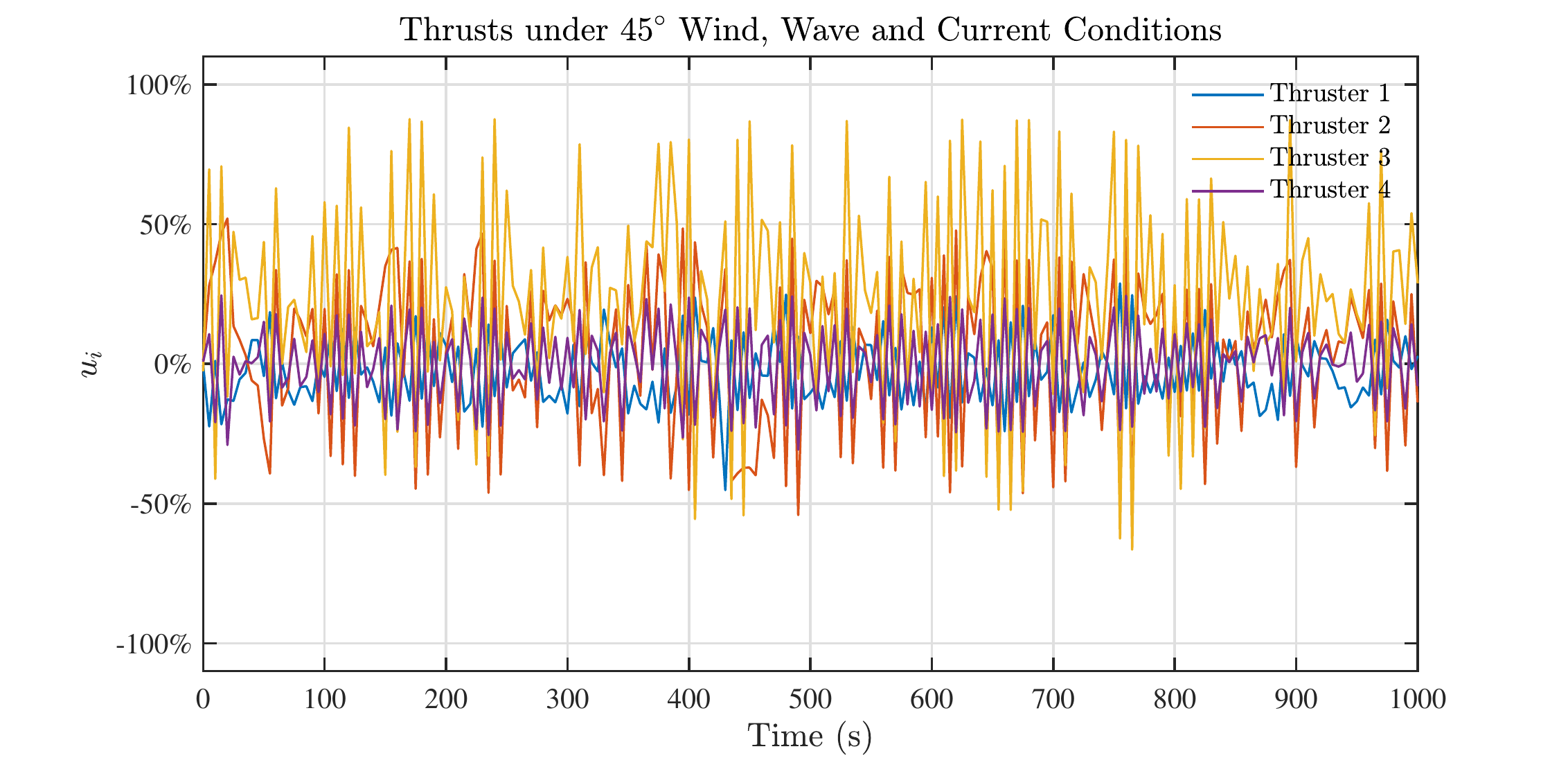}
        \caption{Thrust of Thrusters under Env.1 of 45$^{\circ}$}
        \label{fig:thrust_under_45_thrust}
    \end{figure*}

\begin{figure*}[ht]
        \centering
        \includegraphics[width=0.9\textwidth]{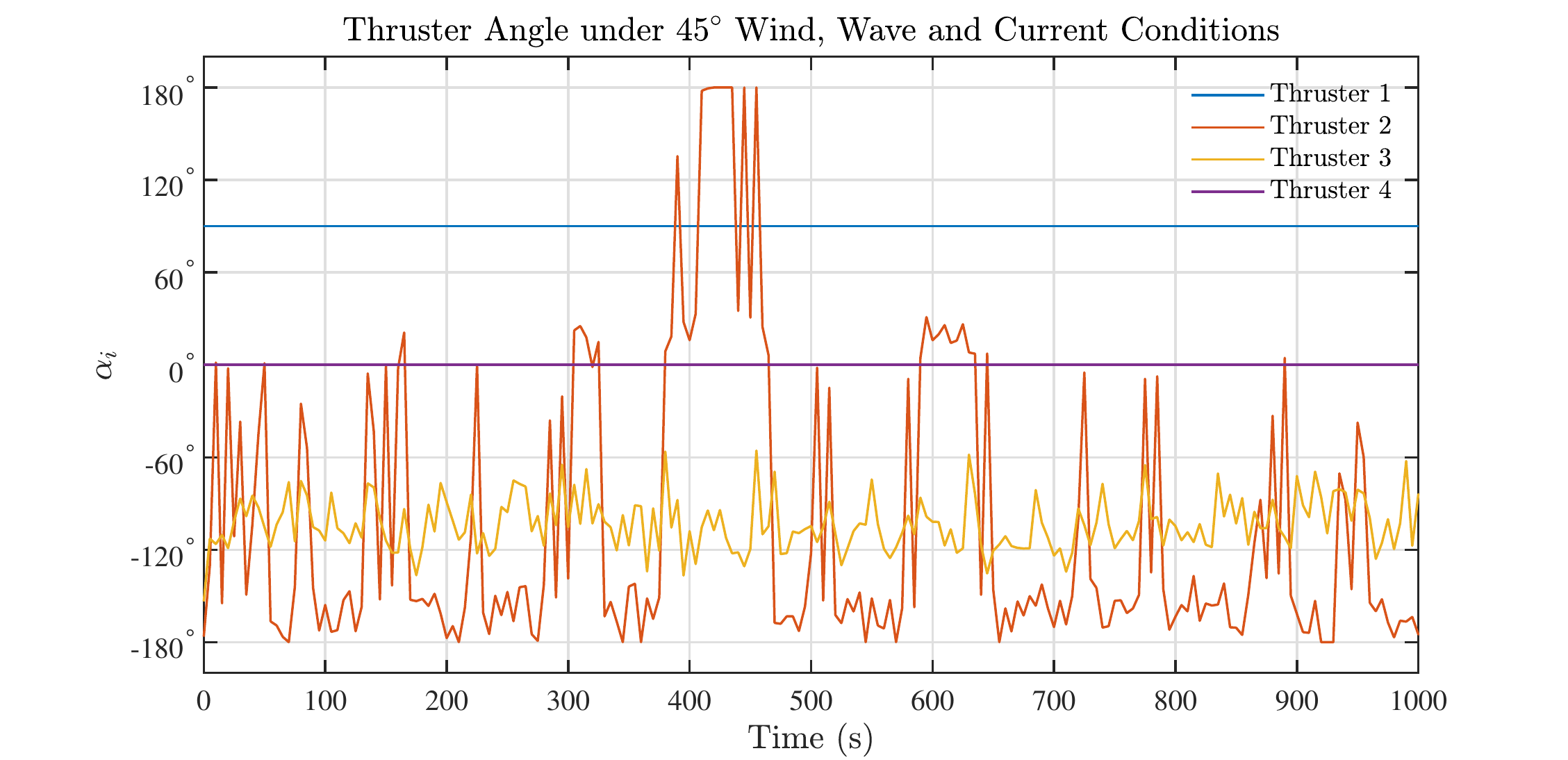}
        \caption{Direction of Thrusters under Env.1 of 45$^{\circ}$}
        \label{fig:thrust_under_45_direction}
    \end{figure*}

    At the same time, shown in Figure \ref{fig:thrust_under_45_thrust}\ref{fig:thrust_under_45_direction} the thrust of most of ships' propellers is around 30\% of the maximum thrust, and the angle of the azimuth thrusters is kept within a certain range, showing a good energy-saving performance.
    
    \FloatBarrier

    \FloatBarrier

    \subsection{4-corner box manoeuvre test}

\begin{figure}[ht]
        \centering
        \subfloat[{Target poses}]{\includegraphics[height=0.3\textwidth]{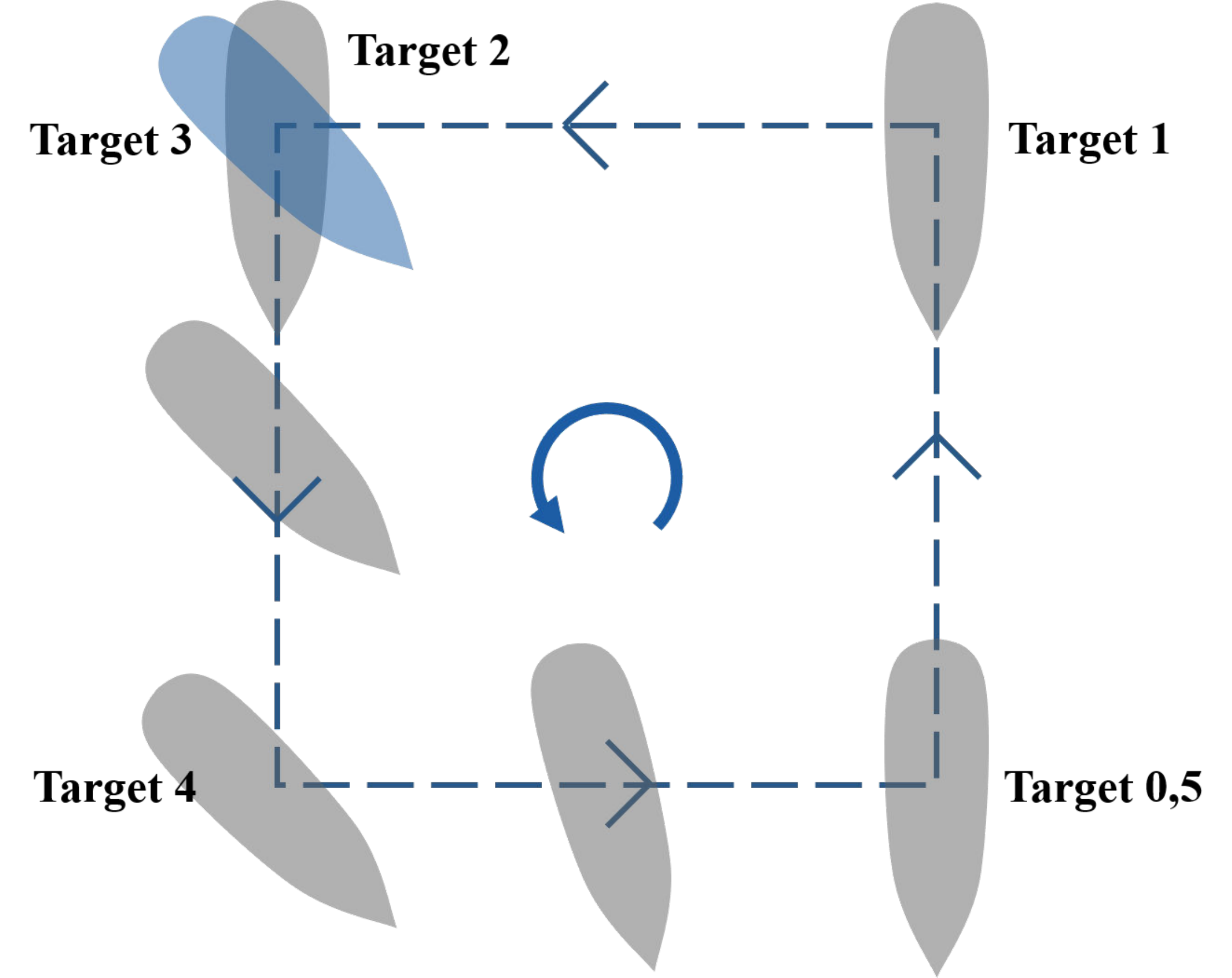}}
        \subfloat[{Time-lapse photography of experiment }]{\includegraphics[height=0.3\textwidth]{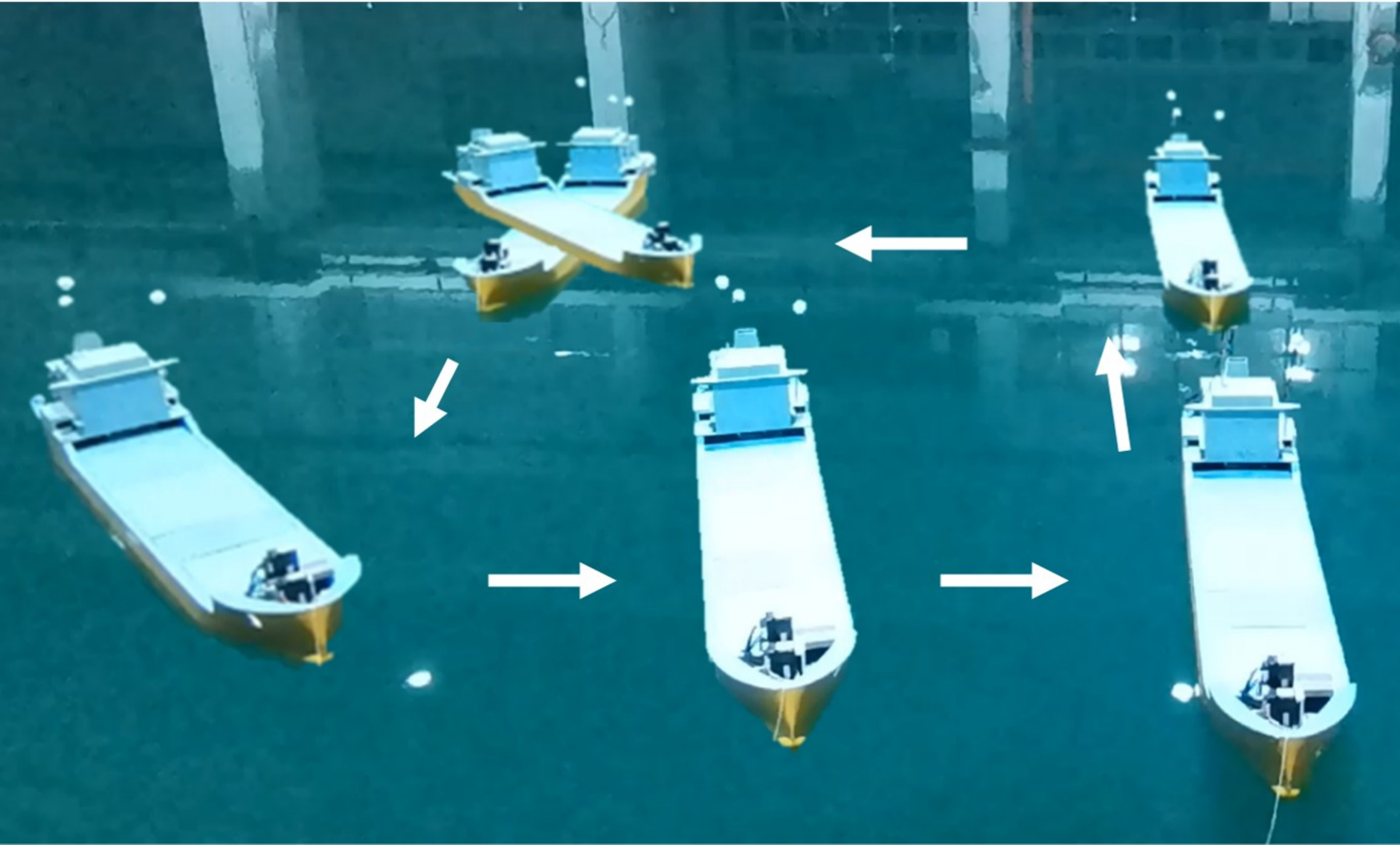}}
        \caption{Schematic plot of the 4-corner box manoeuvre test}
        \label{fig:4_box_target}
    \end{figure}

    On the basis of using simulation to obtain the control parameters shown in Section \ref{sec:Results:pk}, In order to evaluate control effect of the parameters in the model experiment and the ship's motion capability of DP control module, a 4-corner box manoeuvre test was carried out. The test diagram is shown in Figure \ref{fig:4_box_target}. The test is realized by changing the ship target coordinate points to four vertices of a square. Through this test, all motion performance of ship under three degrees of freedom can be detected.

    \FloatBarrier

    \begin{figure}[ht]
        \centering
        \subfloat[{Simulation result}]{\includegraphics[width=0.8\textwidth]{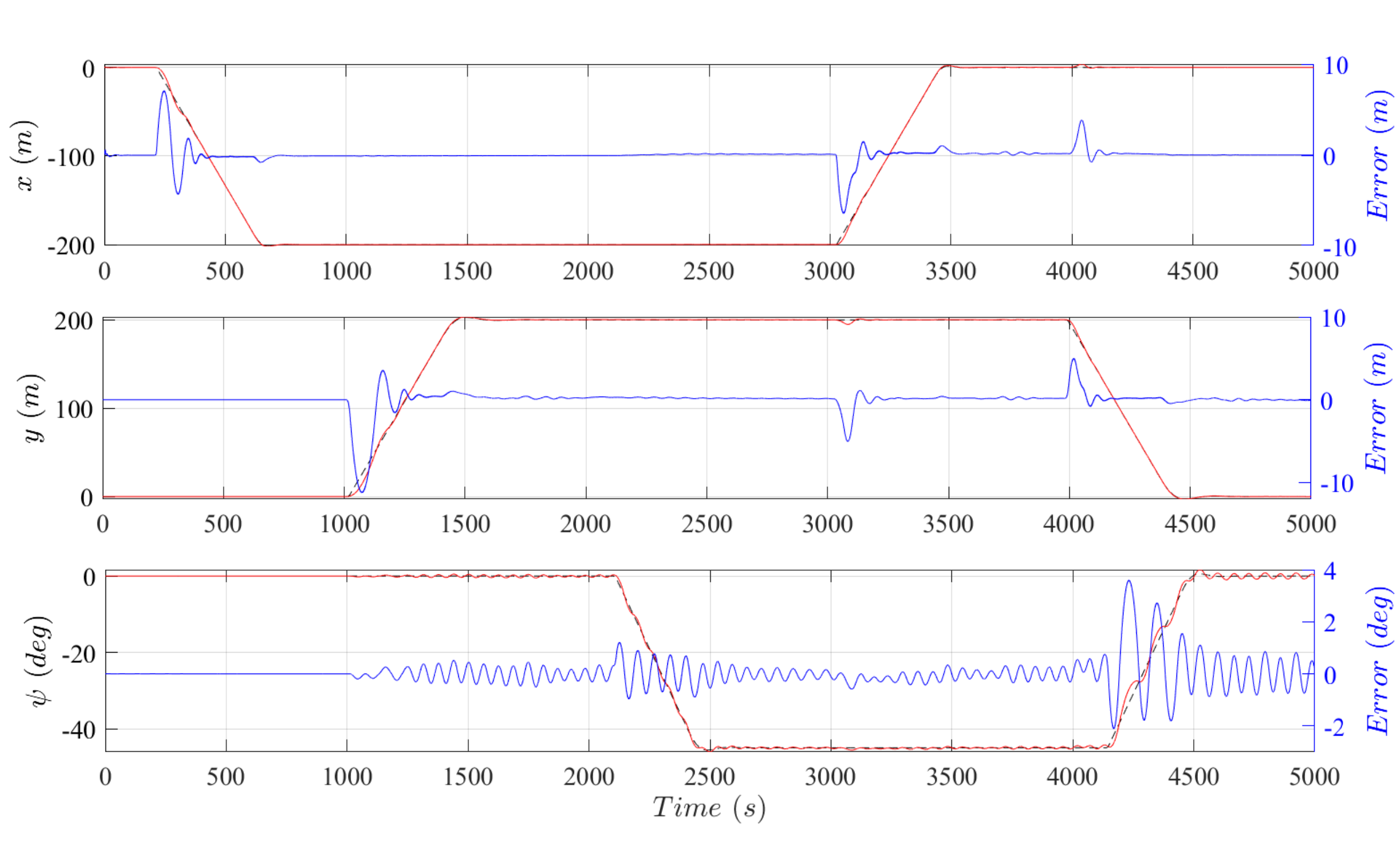}}
        \\
        \subfloat[{Experiment result}]{\includegraphics[width=0.8\textwidth]{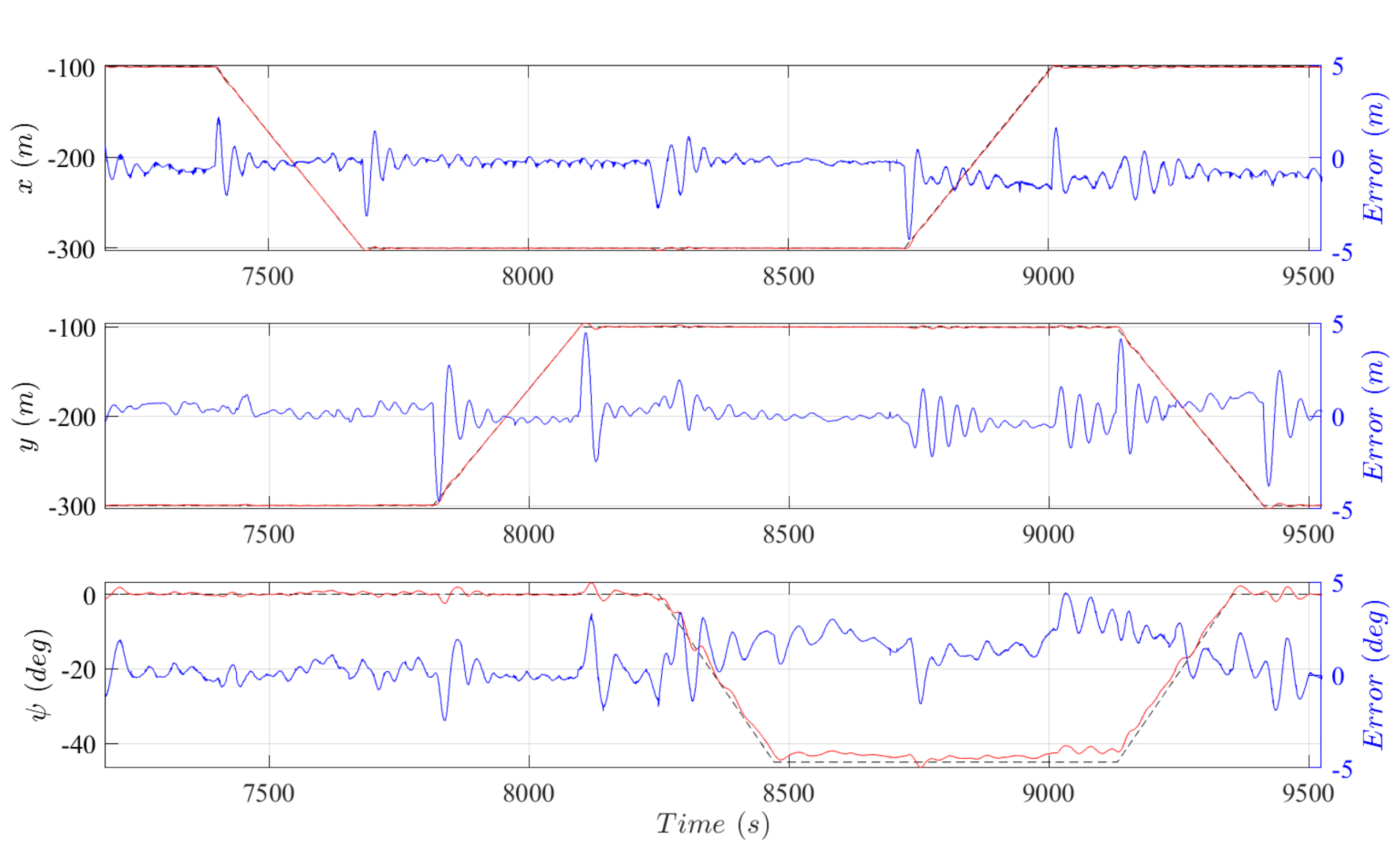}}
        \caption{Time variation of planar motion in 4-corner box manoeuvre test}
        \label{fig:4_box_Time}
    \end{figure}
    \FloatBarrier

\begin{figure}[ht]
        \centering
        \subfloat[{Simulation result}]{\includegraphics[height=0.4\textwidth]{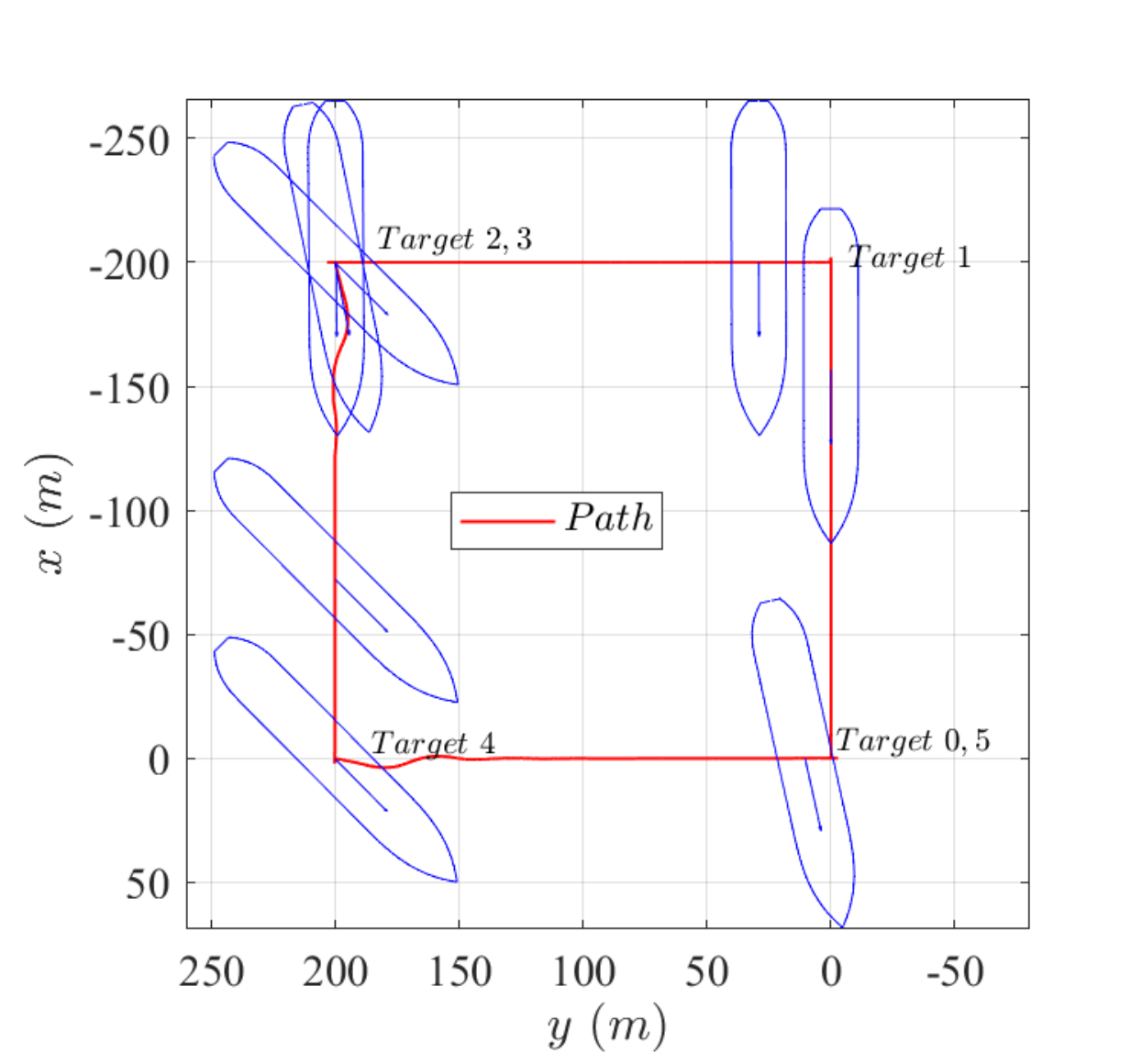}}
        \subfloat[{Experiment result}]{\includegraphics[height=0.4\textwidth]{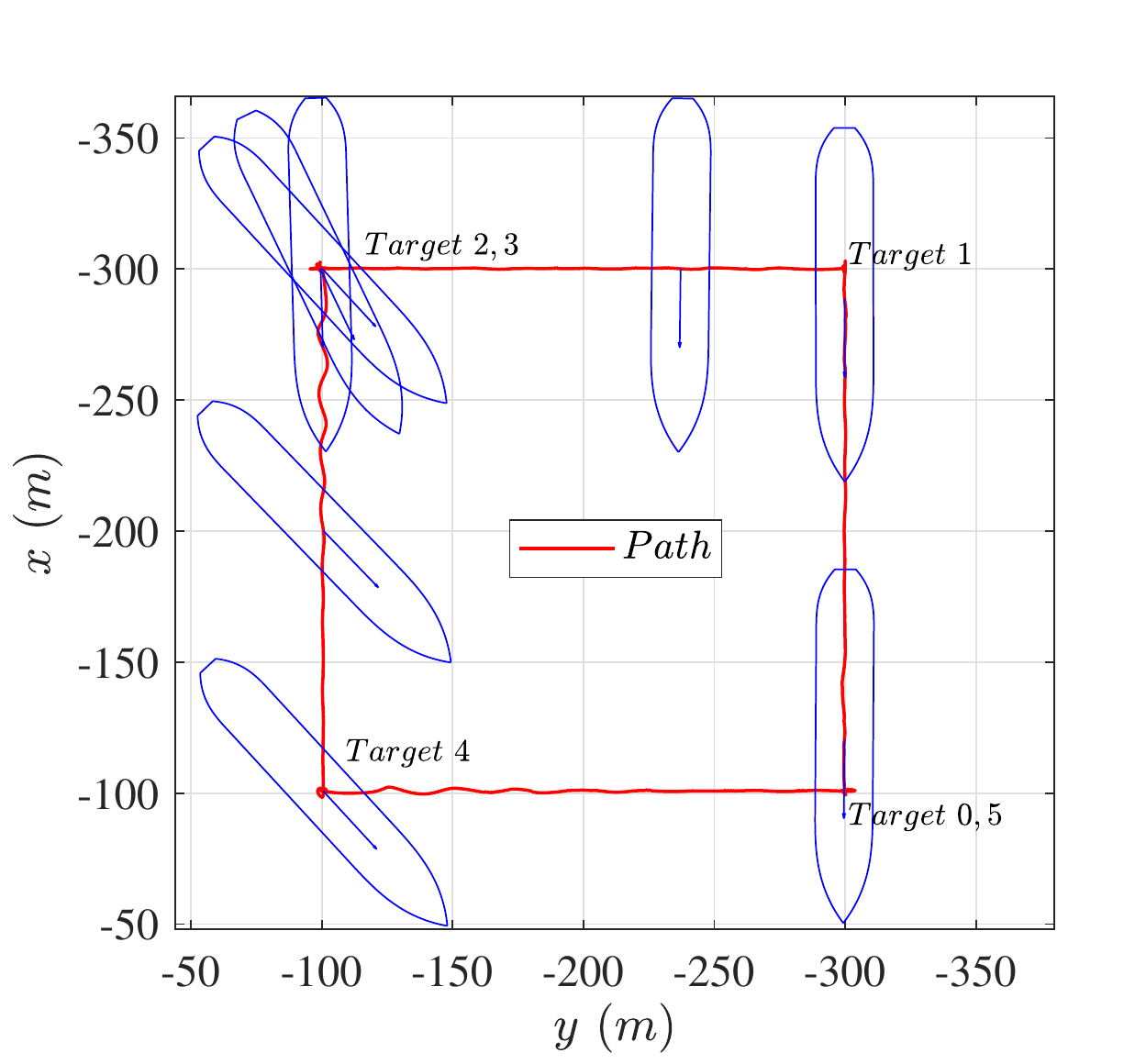}}
        \caption{Trajectory of ST in 4-corner box manoeuvre test}
        \label{fig:4_box_Trajectory}
    \end{figure}

    The movement track of the shuttle tanker model in the simulation and experiment is shown in Figure \ref{fig:4_box_Time}, \ref{fig:4_box_Trajectory}.

    It can be seen that the position error of the ship in both simulation and experiment is maintained within 1m during the course of travel, and only a position error of about 5m to 10m occurs when switching the moving state. The heading angle error is within 5 degrees during the whole process.

    \FloatBarrier

    \section{Conclusion}

    A hybrid simulation and experiment test platform for DP vessels is successfully built and tested. In the test, firstly, the online parameter adjustment is used for parameter tuning in the simulation environment. On this basis, a series of DP capability tests are carried out in the simulation and experiment environment. The results show that the control system is equally effective in physical model experiment and numerical simulation environment.

    Through the above research, some conclusions can be drawn as follows:

    \begin{itemize}
        \item           The accurate simulation model takes account of the fluid memory effect and frequency correlation, which provides strong support for the motion response simulation of DP task and makes the parameter setting more accurate.
        \item          The controller of modular switchable algorithm and on-line parameter adjustment facilitates parameter tuning and can be extended to comparative testing of various algorithms.
        \item          In the scale model experiment, the similarity principle is used to realize the interactive control of the full scale. The control parameters of the full scale simulation can provide effective reference for the model experiment and save the adjustment time and cost of the parameters tuning.
    \end{itemize}

    The future work on simulation and experiment platform construction should include: further refinement of the simulation model, especially the propeller part, so that the friction resistance of the actual ship propeller and the ship wake and thrust deduction can be taken into account in the simulation model. On this basis, the dynamic positioning function of propeller failure can be considered to further improve the dynamic positioning performance of ships in complex operating environment.

    \section*{Acknowledgement}

    This work is supported by the Major Science and Technology Projects of Hainan Province (No. ZDKJ2019001), the High-Tech Ship Research Project Supported by Ministry of Industry and Information Technology (No. MC-202030-H04) and in part by the State Key Laboratory of Ocean Engineering (Shanghai Jiao Tong University) (Grant No. 1915), to which the authors are most grateful.

\bibliographystyle{unsrt}  
\bibliography{templateArxiv}

\end{document}